# Equation of state from symmetry

**Ti-Wei Xue [1] and Zeng-Yuan Guo [1,*]**

[1] Key Laboratory for Thermal Science and Power Engineering of Ministry of Education, Department of Engineering Mechanics, Tsinghua University, Beijing 100084, China

* Correspondence: demgzy@tsinghua.edu.cn

## Abstract

Equation of state (EOS) describes the thermodynamic properties of substances. It has important applications in many fields such as power mechanics, geophysics, astrophysics, and detonation physics [1-10]. Currently, most EOSs have been constructed using the ideal gas EOS as the base framework. However, this is inappropriate for the substances with high pressures or low temperatures that are far from the ideal gas state. Here we extract a concept of "ideal solid" that is symmetrical with ideal gas and propose its EOS, $TS = R_P P$, where $R_P$ is a constant. We verify that the ideal solid EOS represents the thermodynamic properties of the high-pressure and low-temperature limiting state. It indicates that the thermodynamic properties of substances have symmetrical characteristic. Then we develop a universal model by interpolation of the ideal gas and the ideal solid limits. We verify that the universal EOS has high description accuracy in a wide region. The concept of ideal solid and its EOS potentially provides a new direction for the development of the EOS theory.

The study of thermodynamic properties of substances has always been important research topics. It covers a wide range of applications. For example, the cooling performance of an air conditioner depends on the thermodynamic properties of refrigerants [1]. The dynamical evolution of the Earth's interior or celestial bodies is also closely related to the thermodynamic properties of substances [2-9]. The thermodynamic properties of detonation products directly influence the brisance of explosives or munitions [10]. Equation of state (EOS) is the basic theory of the thermodynamic properties of substances. It gives the relationship between various thermodynamic quantities. Studying EOS deepens the understanding of the nature of the motion of substances in nature. EOS is also indispensable for solving a large number of practical problems in engineering technology. In recent years, with the development of experimental technologies, the knowledge about the thermodynamic properties of substances with high pressures or high densities is gradually becoming abundant [11-19], which puts forward further requirements for the development of the EOS theory.

The ideal gas EOS is the earliest quantitative description of the thermodynamic properties of substances and is also the origin of thermodynamics. Its mathematical form is [20, 21]:

$$PV = RT \tag{1}$$

The ideal gas EOS is very general since all substances share this common relationship at the low-pressure and high-temperature limit. A substance that has either very low pressures or high temperatures can be estimated as ideal gas [22]. As the temperature decreases or the pressure increases, actual substance begins to deviate from the ideal gas behavior. To ensure the description accuracy for actual substances, the ideal gas EOS has been modified in many ways. For example, the van der Waals EOS was introduced by adding terms for the intermolecular attraction force and the volume occupied by the molecules themselves to the ideal gas model [20]. The virial EOS was introduced as a Maclaurin series about the ideal gas limit [23]. As the state of substance varies further from the ideal gas state, some complex EOSs containing more parameters have been established. For example, the BWR EOS introduces eight parameters for improving the accuracy in describing high-density substances [24-26]. The usual approach to developing an EOS is to construct formulas based on molecular structural characteristics and statistical thermodynamic theory, and then use experiments to fit the parameters therein [27-30]. These EOSs are generally tedious to calculate and a large amount of experimental data is needed to determine the values of the parameters involved. Therefore, the parameters usually do not have a clear physical meaning, which results that such EOSs have no ability to predict the properties outside their limited ranges of application [31].

In general, most EOSs have been constructed using the ideal gas EOS as the base framework. This is inappropriate for the substances that are far from the ideal gas state. Extensive experience has shown that the thermodynamic behavior of the high-pressure or low-temperature substances differs significantly from that of the low-pressure or high-temperature substances. We need to understand the general laws behind the thermodynamic properties of substances in order to describe the thermodynamic behavior of the high-pressure or low-temperature substances better. Thermodynamics is a science of symmetry [32-34]. Some new parameters and equations in thermodynamics have been discovered with the powerful tool of symmetry[33]. Callen pointed out that thermodynamics is the study of those properties of macroscopic substance that follow from the symmetry properties of physical laws [34]. We infer that the thermodynamic properties of substances have some kind of symmetry. Further, we succeed in finding an EOS that is symmetrical with the ideal gas EOS and it describes the thermodynamic properties of the high-pressure and low-temperature limiting state. The experimental data verifies the validity of this EOS. This provides another base framework for establishing the EOSs of actual substances and brings something new to the fundamental theory of thermodynamics.

## Ideal solid equation of state

When approaching the high-temperature and low-pressure limit, all substances will obey the ideal gas EOS. Considering symmetry, as the thermodynamic state of substance moves away from the ideal gas limit and reach the opposite high-pressure and low-temperature limit, it should obey an EOS symmetric to the ideal gas EOS. In contrast to the concept of ideal gas, we extract a concept of high-pressure and low-temperature limiting state. From the phase diagram, it should be in a solid state, so we call it "ideal solid". Such a state is far from being entirely understood or agreed upon because there is little or no experimental data until now [35, 36]. However, symmetry can tell us something about its thermodynamic properties. For example, in contrast to an ideal gas with infinite specific volume, the specific volume of an ideal solid should tend to zero [37]. Compared to other thermodynamic states, an ideal gas has maximum entropy, while an ideal solid should have minimum entropy [38]. Similar to an ideal gas, an ideal solid does not actually exist but is a common limiting state that various substances tend to as the pressure increases and the temperature decreases.

The symmetry between the ideal gas limit and the ideal solid limit shows that the two physical quantities, temperature and pressure, are symmetric. Then the physical quantities conjugated to temperature and pressure respectively, entropy and volume, should also be a symmetric pair of physical quantities. Based on the symmetry with the ideal gas EOS, we directly infer the ideal solid EOS as follows:

$$TS = R_P P \tag{2}$$

where $R_P$ is the ideal solid constant. Like the ideal gas EOS, the ideal solid EOS has simplicity and generality. It portrays the general thermodynamic behavior of homogeneous substance that does not contain any local information such as phase transitions and critical phenomena.

Thermodynamic deduction shows that the internal energy of an ideal solid depends on pressure only, $U = U(P)$ (see Appendix A). This feature is exactly symmetrical to the fact that the internal energy of an ideal gas depends in temperature only. Thus, by analogy with the specific heat capacity at constant volume, we define a concept of specific work capacity at constant entropy:

$$C_S = \left(\frac{\partial U}{\partial P}\right)_S = -P\left(\frac{\partial V}{\partial P}\right)_S \tag{3}$$

Then the differential of internal energy of an ideal solid becomes $\mathrm{d}U = C_S\,\mathrm{d}P$. Accordingly, the differential of Helmholtz free energy of an ideal solid becomes $\mathrm{d}F = (C_S - R_P)\,\mathrm{d}P$. Thus, the Helmholtz free energy of an ideal solid is also a function of pressure only, $F = F(P)$. By analogy with the specific heat capacity at constant pressure, we further define a concept of specific work capacity at constant temperature:

$$C_T = \left(\frac{\partial F}{\partial P}\right)_T = -P\left(\frac{\partial V}{\partial P}\right)_T \tag{4}$$

An ideal gas satisfies the Mayer formula, $C_P = C_V + R$. A similar parametric relationship exists for an ideal solid, $C_T = C_S - R_P$.

With the help of these two concepts of specific work capacity at constant entropy and specific work capacity at constant temperature, we can make a symmetric statement of thermodynamic stability (see Appendix B). In turn, thermodynamic stability makes the physical meaning of these two concepts more explicit and gives their limitations on values, $C_T > C_S > 0$. This means that the ideal solid constant must be negative, $R_P < 0$. As a result, the entropy in the ideal solid EOS (2) should also be negative. This is not a physical paradox. As an unobservable extensive variable, the positive and negative values of entropy depend on the choice of reference state. That is, there is an implied reference state in the ideal solid EOS (2) that makes the entropy negative. Then its reference state should have maximum entropy, compared to the thermodynamic state described by the ideal solid EOS (2).

In fact, there is an implied reference state of volume in the ideal gas EOS (1). The volume of its reference state is zero, which makes the volume of the thermodynamic state described by the ideal gas EOS (1) always positive. Because volume is an observable variable, this statement seems more like common sense. However, note that the reference state with zero volume is exactly the ideal solid limiting state. That is, the reference state of the ideal gas EOS (1) with respect to volume is the ideal solid state. Then symmetry tells us that the reference state of the ideal solid EOS (2) with respect to entropy should be the ideal gas state. The ideal gas state has maximum entropy, which matches the requirement of the ideal solid EOS (2) for reference state. That is, symmetry also exists for the reference states of the two EOSs (1) and (2).

In engineering practice or experiments, a finite state such as the standard state (STP) or the normal boiling point (NBP) is usually chosen as the reference state for entropy. In order to match these entropy data, an additional constant needs to be introduced to change the reference state of the ideal solid EOS (2). Then the ideal solid EOS becomes:

$$T\,(S - \mathbf{a}) = R_P P \tag{5}$$

where **a** is constant entropy difference between the reference state of the described entropy data and the ideal gas state. Note that the choice of reference state only changes the absolute value of entropy but does not influence its variation characteristics. For equation (5), the internal energy still depends on pressure only.

Moreover, with the concept of specific work capacity at constant entropy, $C_S$, we can get the ideal solid EOS in *P*-*V*-*T* form (See Appendix A):

$$V = -C_T \ln\frac{P}{P_0} - R_P \ln\frac{T}{T_0} + V_0 \tag{6}$$

The subscript 0 denotes a standard state. Compare with the ideal gas *P*-*S*-*T* EOS:

$$S = C_P \ln\frac{T}{T_0} - R \ln\frac{P}{P_0} + S_0 \tag{7}$$

There is also good symmetry between equations (6) and (7).

Some empirical equations corroborate the ideal solid EOSs. The ideal solid EOS, in turn, provides a theoretical support for these empirical equations. For example, the experiments for solid and liquid alkali metals show that the Grüneisen parameter $\gamma$ at high pressures satisfies a simple linear relationship with volume [39]:

$$\frac{\gamma}{V} = \frac{\gamma_0}{V_0} = \text{Const} \tag{8}$$

It was later found that this relationship could be extrapolated to very high pressures with *ab initio* simulations [40]. The Grüneisen parameter is defined in thermodynamics as:

$$\gamma = V\left(\frac{\partial P}{\partial U}\right)_V \tag{9}$$

An ideal solid requires:

$$\frac{\gamma}{V} = \left(\frac{\partial P}{\partial U}\right)_V = \frac{\mathrm{d}P}{\mathrm{d}U} = \frac{1}{C_S} \tag{10}$$

Thus, the constant in equation (8) is just the reciprocal of specific work capacity at constant entropy. The Tait equation is one of classical EOSs describing substances with high pressures [28, 41]. Its mathematical form is:

$$V = -D\ln\frac{P+E}{P_0+E} + V_0 \tag{11}$$

It is an isothermal equation. $D$ and $E$ are the parameters under isothermal condition. In the high-pressure limit, it will be reduced to:

$$V = -D\ln\frac{P}{P_0} + V_0 \tag{12}$$

It is exactly the ideal solid *P-V-T* EOS (6) under isothermal condition. The parameter $D$ is just the specific work capacity at constant temperature. Equation (6)/(12) states that the volume of high-pressure substances decreases linearly with the increase of the logarithmic pressure at constant temperatures. The ultrahigh-pressure data of Al and Cu demonstrates this feature very well (see Fig. 1).

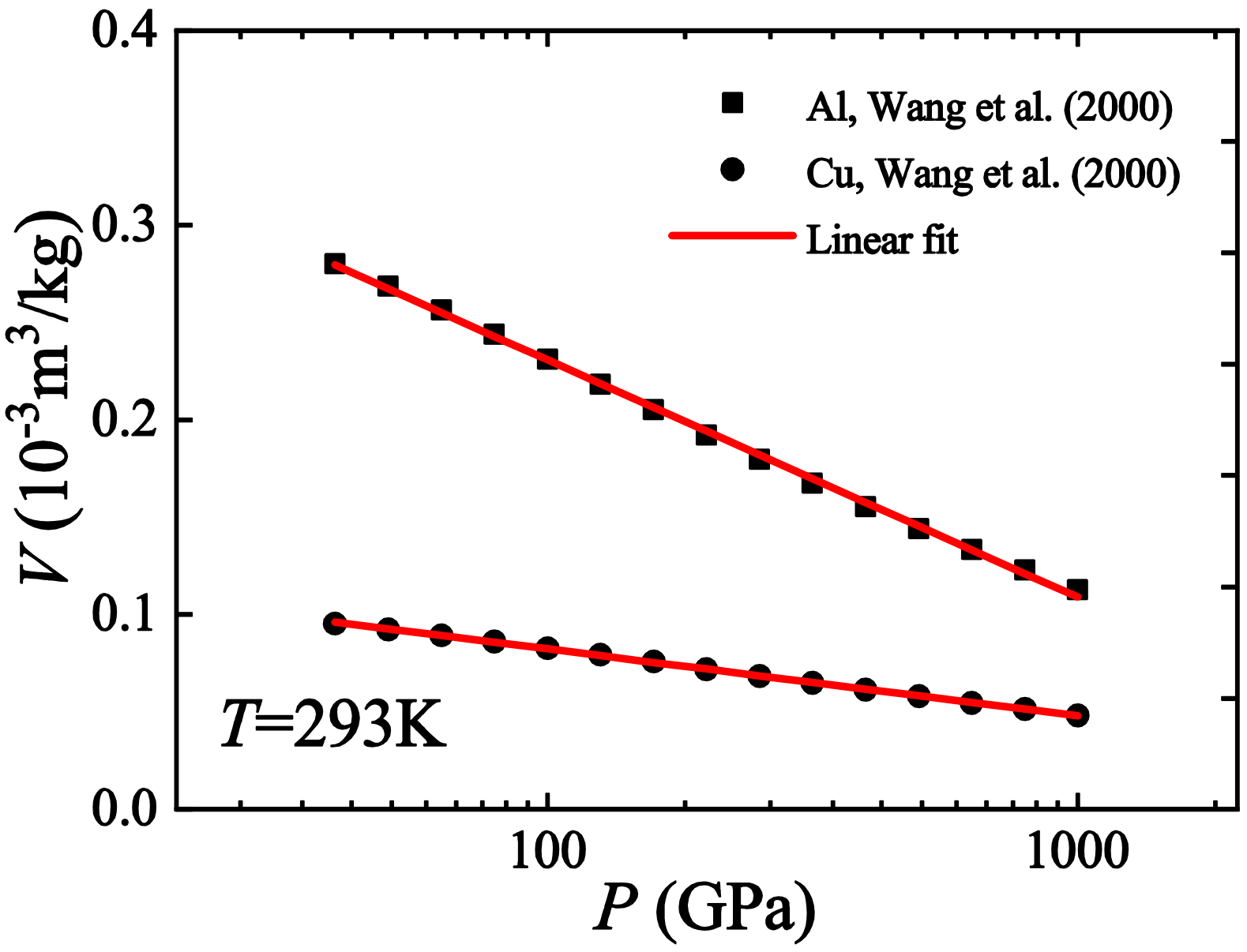

**Fig. 1. Volume variation of ultrahigh-pressure Al and Cu at 293K.** The horizontal coordinate is the logarithmic coordinate. The data comes from Ref. [27]. The volume shows good linearity with the logarithm of pressure, which supports equation (6)/(12).

Next, we verify the ideal solid EOSs (5) and (6) more completely using nitrogen data with relatively high pressures and low temperatures. The data comes from the National Institute of Standards and Technology (NIST) database, where the reference state with respect to entropy is the normal boiling point (NBP). It shows that the selected data has a good agreement with the ideal solid EOSs (see Fig. 2). In addition, the lower the temperature and the higher the pressure, the more accurately the ideal solid EOSs describe. The nitrogen data verifies the validity of the ideal solid model.

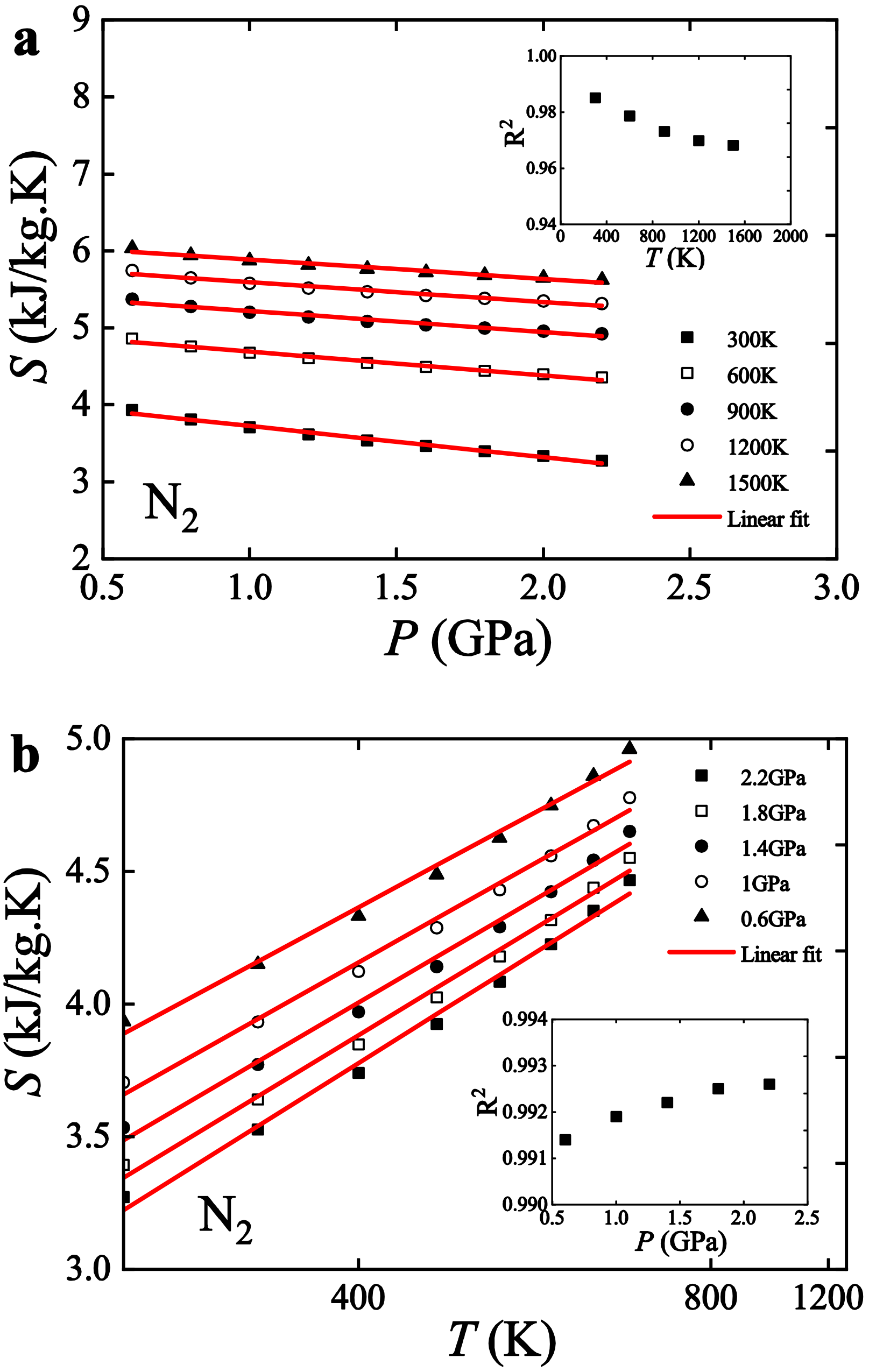

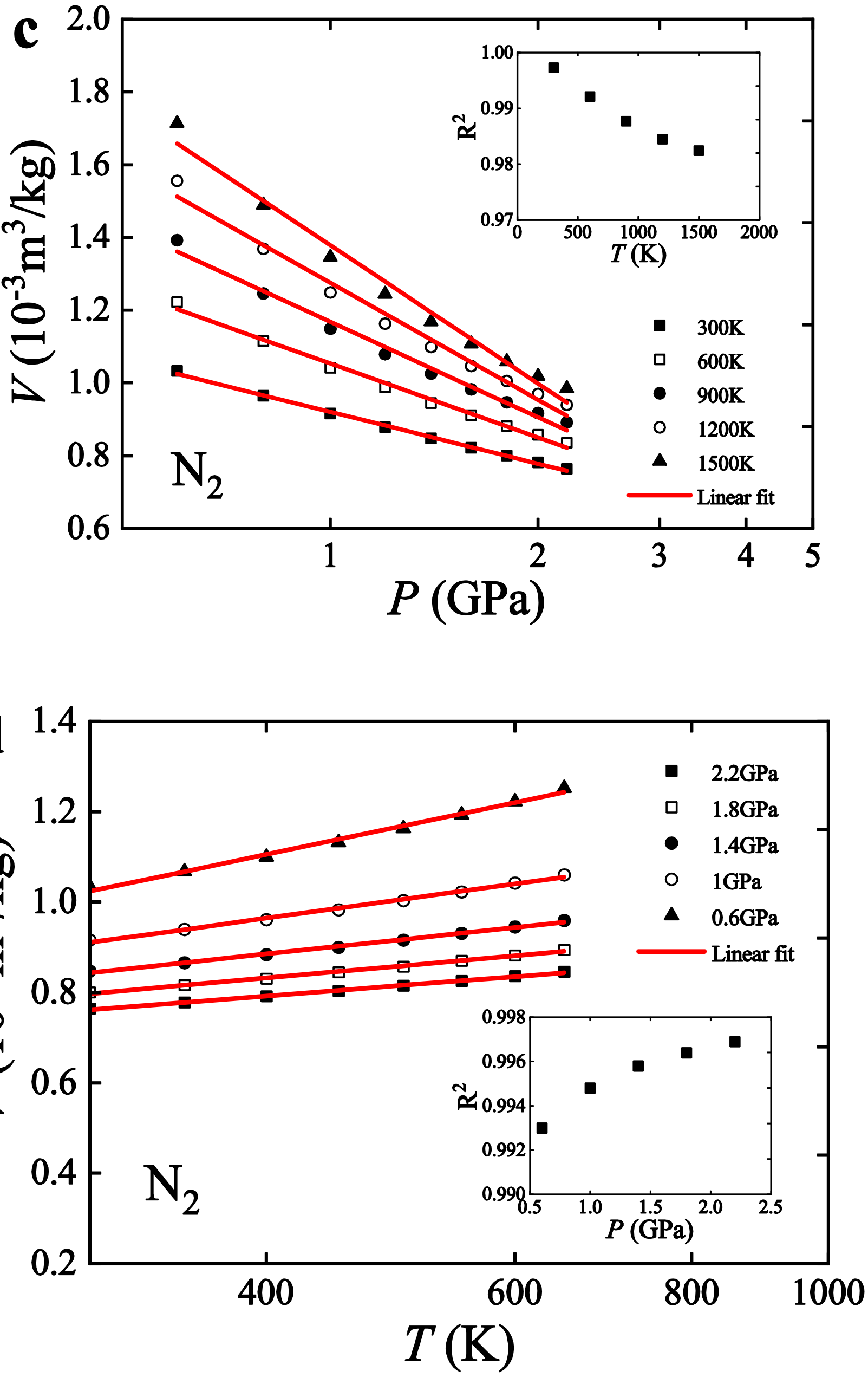

**Fig. 2. Validation of the ideal solid EOSs using nitrogen data.** The insets show their fitting accuracy. **(a)** Entropy versus pressure at different temperatures; **(b)** Entropy versus temperature at different pressures (the horizontal coordinate is the reciprocal coordinate); **(c)** Volume versus pressure at different temperatures (the horizontal coordinate is the logarithmic coordinate); **(d)** Volume versus temperature at different pressures (the horizontal coordinate is the logarithmic coordinate). The ideal solid EOSs predict that substances with higher pressures or lower temperatures will demonstrate more prominently the following thermodynamic characteristics: the entropy of isothermal processes has a linear relationship with pressure and the entropy

of isobaric processes has a linear relationship with the reciprocal of temperature; the volume of isothermal processes has a linear relationship with the logarithm of pressure and the volume of isobaric processes has a linear relationship with the logarithm of temperature. The results show that the nitrogen data has a good agreement with the above thermodynamic characteristics.

The ideal gas constant is an important physical constant related to the thermodynamic properties of substances. It is related only to molar mass. The more the molecules per unit mass, the larger the ideal gas constant. The ideal solid constant is another physical constant related to the thermodynamic properties of substances. Determining the value of the ideal solid constant and its physical characteristics can help to fundamentally understand the laws of thermodynamic properties of substances. Although it is difficult to reach or approach the high-pressure and low-temperature limiting state experimentally, we can predict the value of the ideal solid constant from the available property data.

The slope of the curves in Fig. 2d is:

$$\kappa = \left( \frac{\partial V}{\partial \ln T} \right)_P \tag{13}$$

According to the ideal solid EOS (6), the ideal solid constant is then equal to the opposite of the curve slope in the ideal solid limit:

$$R_P = -\left(\kappa\right)_{\text{i. s.}} \tag{14}$$

The subscript i.s. denotes the ideal solid state. We further find that the slopes of curves at different pressures in Fig. 2d have an excellent linear relationship with the reciprocal of pressure (see Fig. 3), which ensures the availability of prediction of the ideal solid constant. The intercept between the curve and the vertical coordinate in Fig. 3 is just the predicted value of the ideal solid constant.

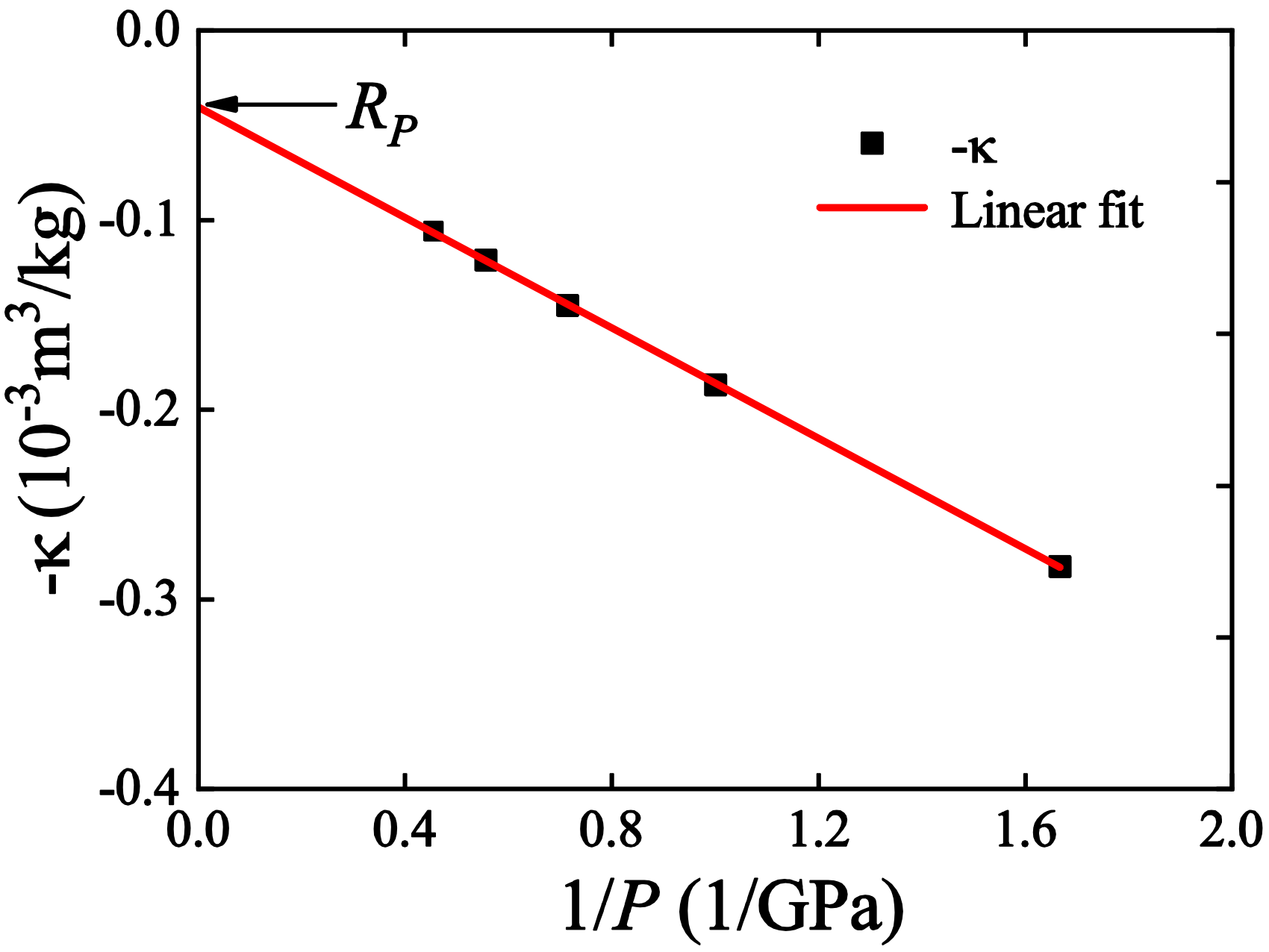

**Fig. 3. Negative slopes of curves at different pressures in Fig. 2d.** The intercept between the curve and the vertical coordinate denotes the negative slope in the high-pressure limit. This is the predicted value of the ideal solid constant we are looking for.

With this method, we have obtained the predicted values of the ideal solid constants for several different substances (see Table 1). Different from the ideal gas constant, the ideal solid constant is mainly influenced by atomic mass. The ideal solid constant has approximately linear relationship with the reciprocal of average atomic mass, $M_T$ (see Fig. 4). The more the atoms per unit mass, the smaller the ideal solid constant. The thermodynamic properties of substances near the ideal solid state should be closely related to the atom state.

**Table 1 Predicted values of the ideal solid constants for several common substances**

| Substance | $H_2O$ | $N_2$ | $CO_2$ | Ne | Ar | Xe |
|---|---|---|---|---|---|---|
| $R_P$ ($10^{-3}$m$^3$/kg) | -0.0997 | -0.0517 | -0.0378 | -0.0157 | -0.0226 | -0.0143 |

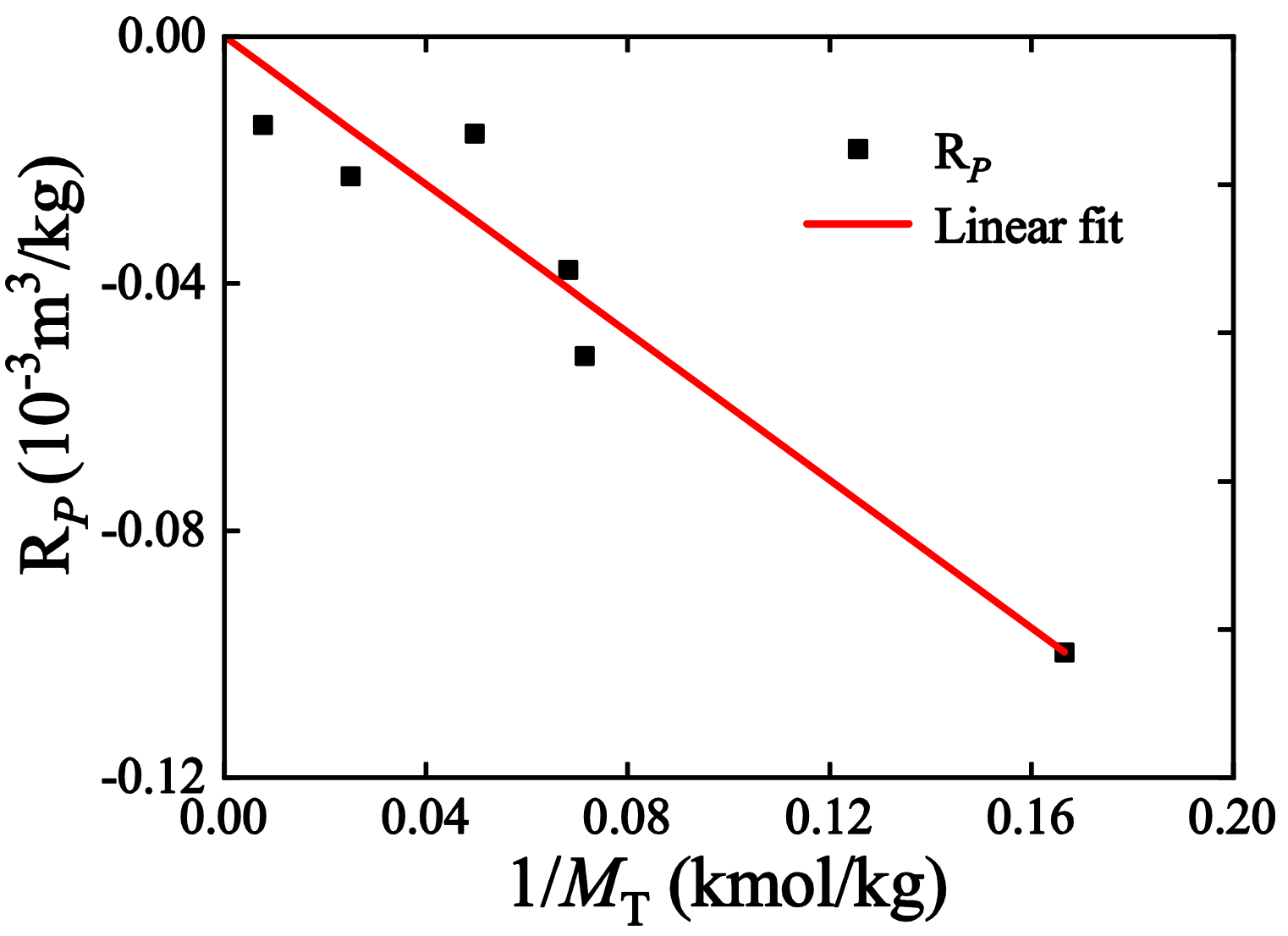

**Fig. 4. Relation between the ideal solid constant and the average atomic mass for different substances.** It shows that the ideal solid constant is closely related to atomic mass.

## Universal equation of state

Either of the ideal gas EOS or the ideal solid EOS is absolutely accurate only for its corresponding limiting state. As substances change from one limiting state to the other limiting state, their thermodynamic properties will become less like those of the initial limiting state and more like those of the other limiting state. The thermodynamic state of actual substance is most likely some kind of superposition of the two limiting states. Next, we present a universal model that is a fusion of the two limiting states for describing the thermodynamic state in the middle region.

The development of the universal model was inspired by Planck's interpolation idea for constructing Planck's law [42, 43]. Planck saw that there were two limiting cases for radiation phenomenon which corresponded to the two thermodynamic relationships: $d^2S/dU^2 = a/U$ in the short-wave limit and $d^2S/dU^2 = b/U^2$ in the long-wave limit. Planck then developed his radiation law, $U = \beta/(e^{-\beta/\alpha T} - 1)$, by interpolating between these two limiting relationships. Here is a similar situation. The first order derivatives of the two limiting thermodynamic states have exponents of $-1$ and $-2$ (See Appendix C). There are four such relationship pairs in a two-degree-of-freedom thermodynamic system. Interpolation between these extremes leads to two following universal EOSs (See Appendix C):

$$V = R'\left(\frac{T}{P} - \frac{T_0}{P_0}\right) - R_P'\ln\frac{T}{T_0} - C_T'\ln\frac{P}{P_0} + V_0 \tag{15}$$

$$S = R_P'\left(\frac{P}{T} - \frac{P_0}{T_0}\right) - R'\ln\frac{P}{P_0} + C_P'\ln\frac{T}{T_0} + S_0 \tag{16}$$

Note that there is a difference in the physical meanings of the parameters before and after interpolation. We call the parameters after interpolation universal parameters and use the superscript ' for the sake of distinction. When approaching the ideal gas limit, the universal EOSs degenerate to the ideal gas EOSs; when approaching the ideal solid limit, degenerate to the ideal solid EOSs. Since both the limiting models can only be applied to homogeneous substances not located in regions with rapid property variations such as near the critical point or the two-phase region, then the universal model is not expected to apply in those regions either.

With temperature and pressure as independent variables, the universal model gives the differential of internal energy:

$$\mathrm{d}U = \left(C_P' - R'\right)\mathrm{d}T + \left(C_T' + R_P'\right)\mathrm{d}P \tag{17}$$

Define the universal specific heat at constant volume, $C_V'$ and the universal specific work at constant entropy, $C_S'$:

$$\begin{cases} C_V' = \left(\dfrac{\partial U}{\partial T}\right)_P \\ C_S' = \left(\dfrac{\partial U}{\partial P}\right)_T \end{cases} \tag{18}$$

Then there are two parameter relations at the same time for the universal model:

$$\begin{cases} C_P' = C_V' + R' \\ C_T' = C_S' - R_P' \end{cases} \tag{19}$$

The parameters before and after interpolation have the following relationships under the universal model:

$$\begin{cases} C_V = C_V' + C_S'\left(\dfrac{\partial P}{\partial T}\right)_V \\ C_P = C_P' - R_P'\dfrac{P}{T} \end{cases};\quad \begin{cases} C_S = C_S' + C_V'\left(\dfrac{\partial T}{\partial P}\right)_S \\ C_T = C_T' + R'\dfrac{T}{P} \end{cases} \tag{20}$$

When approaching the two limits, the universal parameters can again degenerate to the corresponding conventional ones:

$$\begin{cases}\left(C_V = C_V{}'\right)_{\text{i. g.}} \\ \left(C_P = C_P{}'\right)_{\text{i. g.}}\end{cases}; \quad \begin{cases}\left(C_S = C_S{}'\right)_{\text{i. s.}} \\ \left(C_T = C_T{}'\right)_{\text{i. s.}}\end{cases} \tag{21}$$

The subscript i.g. denotes the ideal gas state.

The universal parameters have explicit macro definitions or calculation formulas, so their physical meanings are also clear. We can a priori calculate the values of the universal parameters for each thermodynamic state. We use nitrogen data to compare the universal specific heat at constant pressure, $C_P{}'$, and the universal specific work at constant temperature, $C_T{}'$, with their conventional counterparts (see Fig. 5). Results shows that $C_P{}'$ and $C_T{}'$ have better nature of constant, compared to $C_P$ and $C_T$, respectively. This indicates that the universal model has wider application range. More than that, the universal model also predicts the general physical characteristics of the traditional parameters. For example, from equation (20), the specific heat at constant pressure, $C_P$, has linear relation with pressure under isothermal condition and the specific work at constant temperature, $C_T$, has linear relation with the reciprocal of pressure. These characteristics are consistent with nitrogen data (see Fig. 5). Go back to the slope of curves in Fig. 2d. Under the universal model, its expression becomes:

$$\kappa = \left(\frac{\partial V}{\partial \ln T}\right)_P = T\left(\frac{\partial V}{\partial T}\right)_P = -R_P{}' + R'\frac{T}{P} \tag{22}$$

which nicely characterizes the data in Fig. 3.

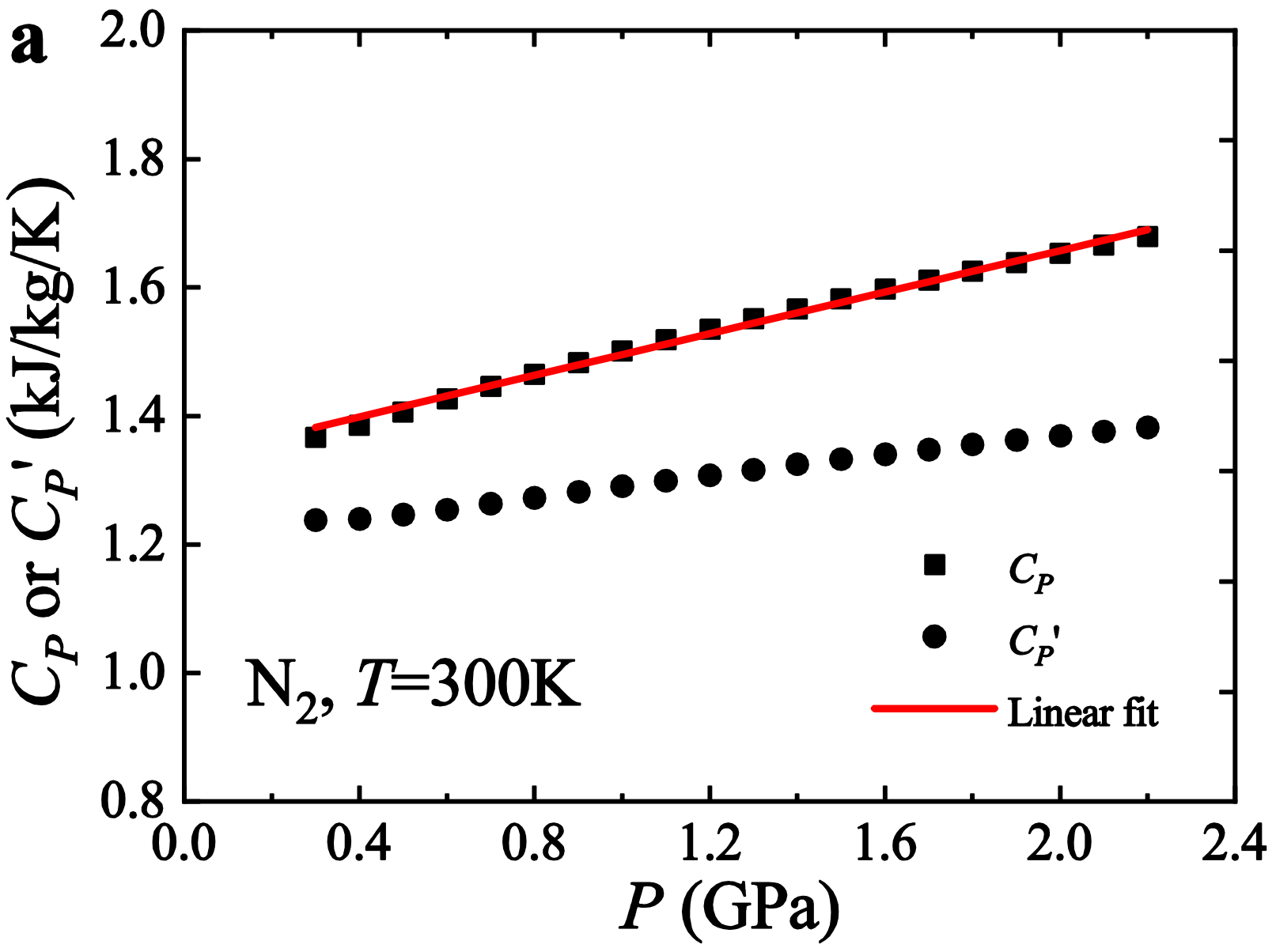

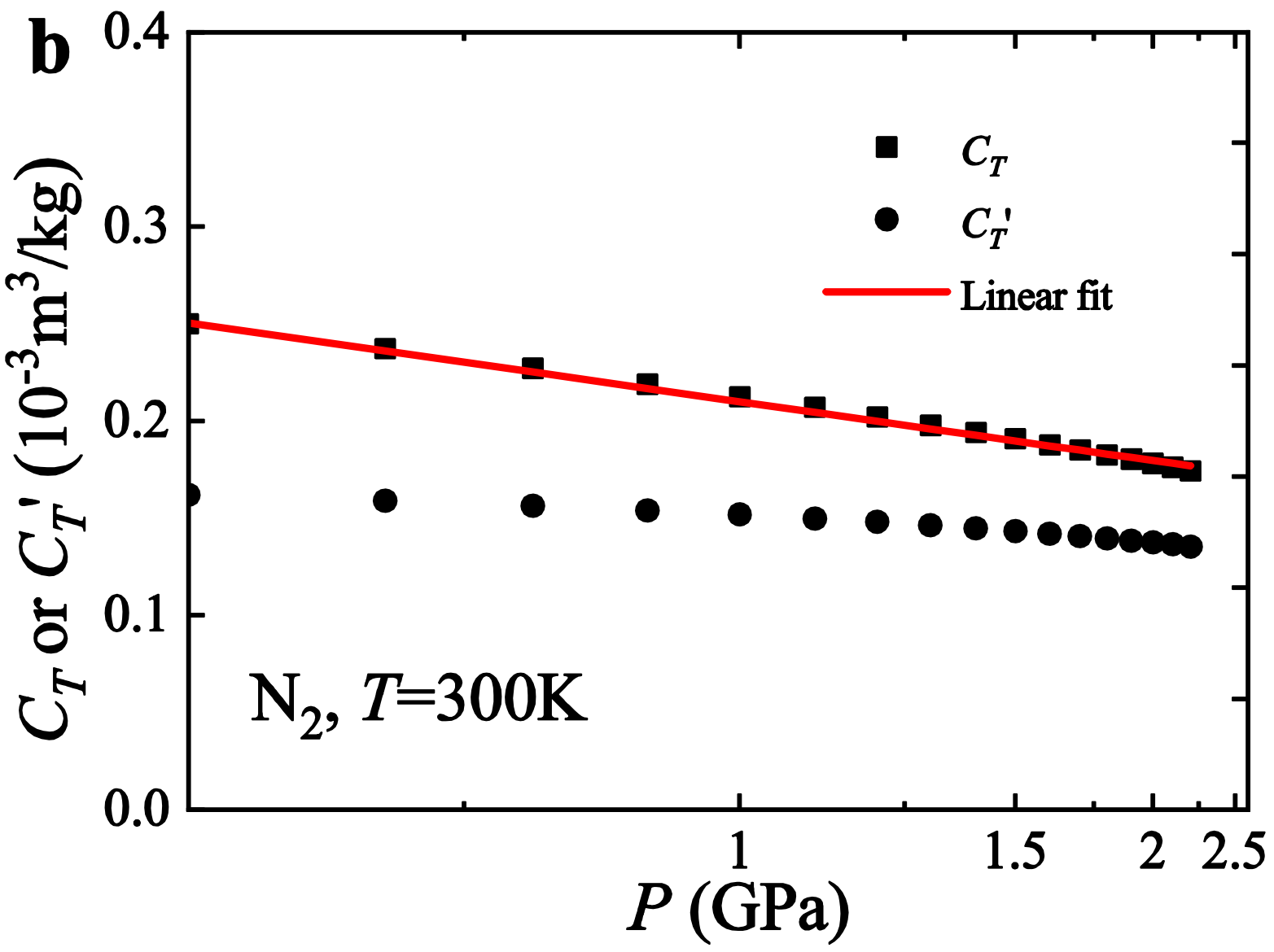

**Fig. 5. Comparison of parameters before and after interpolation using nitrogen data**. (**a**) $C_P$ and $C_P'$ versus pressure at 300K; (**b**) $C_T$ and $C_T'$ versus pressure at 300K (the horizontal coordinate is the reciprocal coordinate). Results show that the universal parameters have better nature of constant. Moreover, the higher the pressure, the closer $C_T$ is to $C_T'$; while the lower the pressure, the closer $C_P$ is to $C_P'$. $C_P$ has linear relation with pressure under isothermal condition; while $C_T$ has linear relation with the reciprocal of pressure. All these features are consistent with equation (20).

The following is a more complete verification for the universal EOSs using helium-4 data from NIST and compare the universal EOSs with the ideal gas EOSs and the ideal solid EOSs at the same time (see Fig. 6). Results shows that the ideal gas model has better performance in high-temperature or low-pressure regions, while the ideal solid model has better performance in high-pressure or low-temperature regions. The entropy and the volume predicted by the ideal gas model are always larger than actual data, while the entropy and the volume predicted by the ideal solid model are always smaller than actual data. The ideal gas curves of entropy and volume have the greater curvatures; while the ideal solid curves of entropy and volume have the smaller curvatures. The ideal gas and the ideal solid models give the upper and lower limits of a thermodynamic state. Compared to the two limiting models, the universal model excellently reproduces the whole property data. Actual substances behave with the nature of superposition of the two limiting states. The universal model reveals the respective proportions of the contribution from the two limiting states.

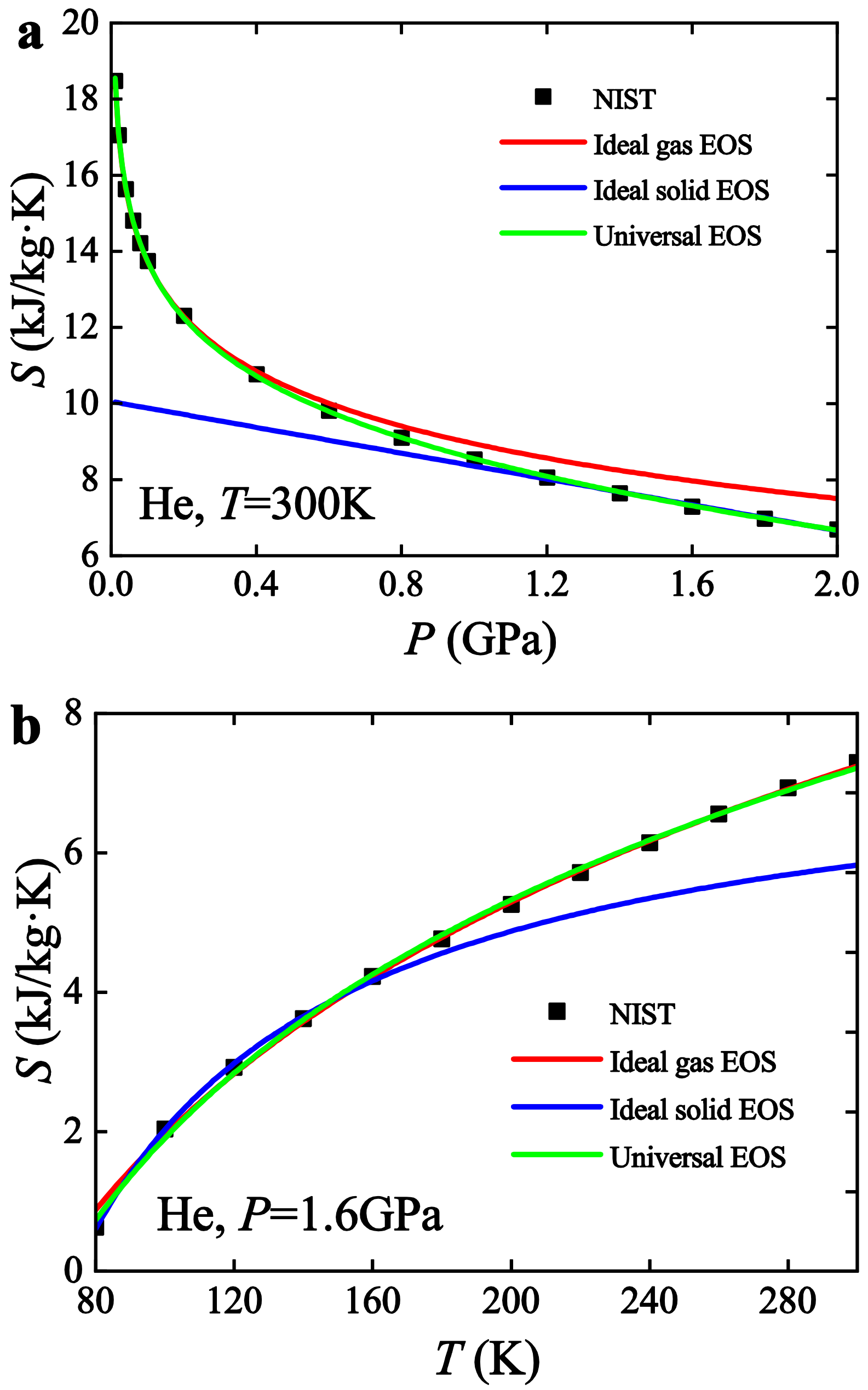

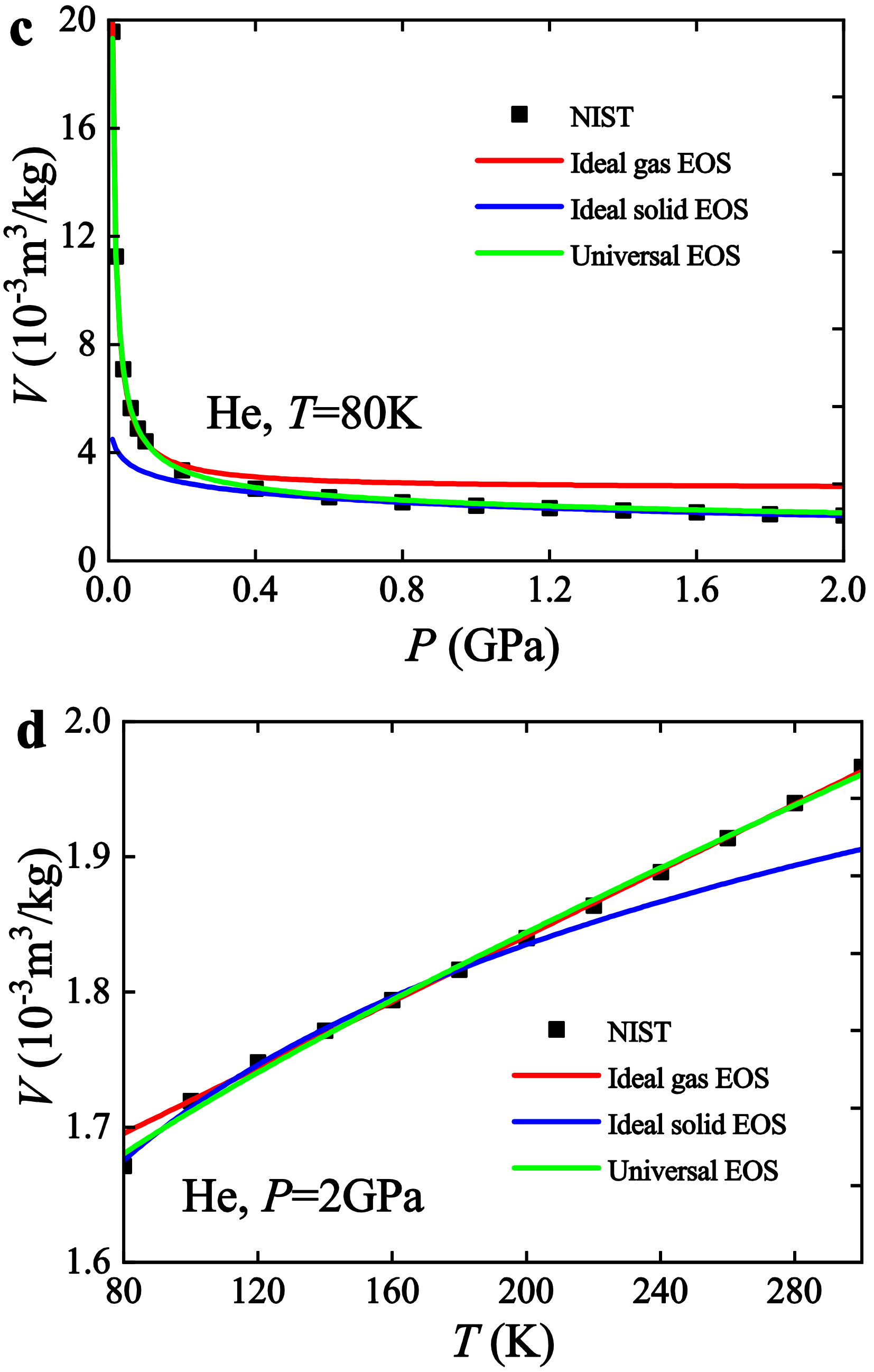

**Fig. 6. Validation of the universal EOSs using helium-4 data and comparison with the ideal gas EOSs and the ideal solid EOSs. (a)** Entropy versus pressure at 300K; **(b)** Entropy versus temperature at 1.6GPa; **(c)** Volume versus pressure at 80K; **(d)** Volume versus temperature at 2GPa. Here, we are more concerned with the variation characteristics of entropy and volume with pressure and temperature, i.e., their derivative properties. Because we have to use equation (5) as the ideal solid *P*-*S*-*T* EOS, for the sake of fairness here we use the Clausius equation, $P\,(V - \mathbf{b}) = RT$, as the ideal gas *P*-*V*-*T* EOS. It has the same derivative properties for volume as equation (1). It can be seen that the ideal gas model is more consistent with the

property data in high-temperature or low-pressure region, while the ideal solid model is more consistent with the property data in high-pressure or low-temperature region. The universal model has good performance in the whole wide region, which verifies the validity of the interpolation pattern.

We use nitrogen data to further compare the universal *P-V-T* EOS (15) with the second-order virial EOS and the BWR EOS (see Fig. 7). As with the universal *P-V-T* EOS, the second-order virial EOS has three parameters. However, the second-order virial EOS is applicable only in lower pressure region. The BWR EOS with eight parameters has more complex form. In return, it broadens the application in high pressure region. However, as the pressure increases further, its description becomes less accurate. Compared with the BWR EOS, the universal *P-V-T* EOS not only has fewer parameters and a simpler structure, but also has a wider application range and a higher descriptive accuracy especially for high pressure region. Another advantage of the universal *P-V-T* EOS is that its parameters have clear macro expressions and can be calculated from other prior knowledge by thermodynamic deduction. The virial coefficients and the parameters in the BWR EOS are empirical parameters and are more often determined by fitting experimental data. Currently, most of EOSs including the virial EOS and the BWR EOS are constructed using the ideal gas EOS as the basic framework. No matter how complex these EOSs are, their common feature is that the descriptive accuracy will decrease as the pressure increases or the temperature decreases. The universal EOS, however, uses both the limiting EOSs, i.e., the ideal gas EOS and the ideal solid EOS, as the basic framework and thus demonstrates significant advantages.

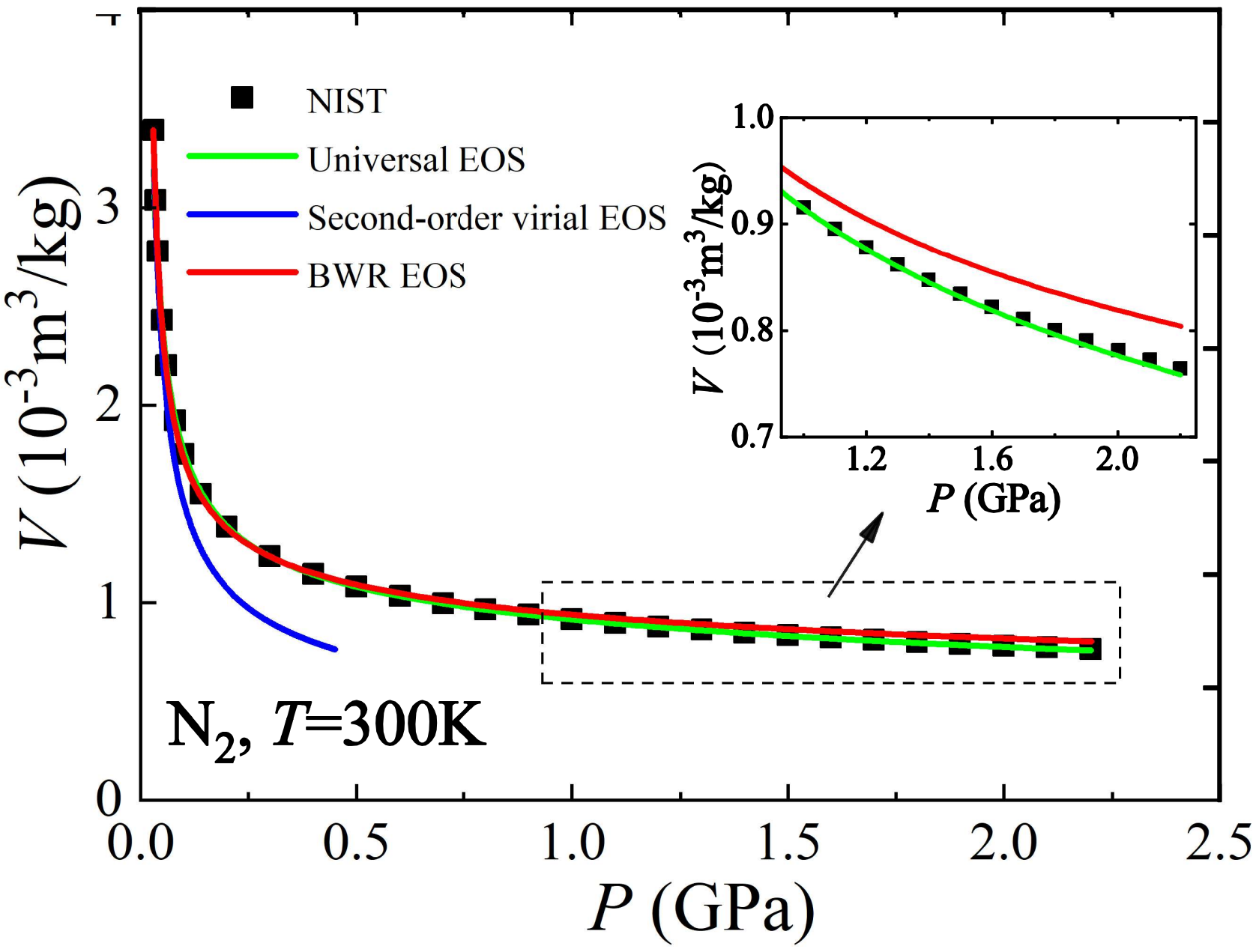

**Fig. 7. Comparison of the universal *P-V-T* EOS with the second-order virial EOS and the BWR EOS using nitrogen data.** The values of the parameters in the second-order virial EOS and the BWR EOS come from Refs. [44] and [26], respectively. It can be seen that the universal *P-V-T* EOS has a wider application range and a higher descriptive accuracy for high pressure region.

In summary, we propose an ideal solid EOS that is symmetrical with the ideal gas EOS. We verify that the thermodynamic properties of substance have symmetrical characteristic. We further develop a universal model, inspired by Planck's interpolation idea. EOS is a theory of the macroscopic thermodynamic properties of substance. In this work, we adopt a symmetry-based, completely macroscopic approach. Therefore, the result should be highly reliable. This may provide a new perspective for the development of the EOS theory.

EOS in a narrow sense usually refers to *P-V-T* EOS. However, only one *P-V-T* EOS does not cover all the thermodynamic properties of a substance. *P-S-T* EOS is also needed and should have the same status as *P-V-T* EOS. For example, the universal model gives the *P-V-T* EOS (15) and the *P-S-T* EOS (16), simultaneously. The two EOSs constitute a complete description of thermodynamic properties of substance. When both equations are present at the same time, some new hidden properties can be discovered. For example, we get a new EOS by combining equations (15) and (16):

$$C_S'S+C_V'V=C_S'R_P'\frac{P}{T}+C_V'R'\frac{T}{P}+\left(C_T'C_V'+C_S'R'\right)\ln\frac{T}{P}+\mathbf{c} \tag{23}$$

where **c** is a constant. Equation (23) is a universal EOS containing three variables, the volume, the entropy and the ratio of temperature to pressure, $\frac{T}{P}$, where $\frac{T}{P}$ should be viewed as a combined physical quantity. Equation (23) can be used to predict the irreversibility of shock adiabatic processes (Hugoniots). The ratio of temperature to pressure of some Hugoniots approximately remains constant [45-48] and thus according to equation (23), entropy is linearly inversely proportional to volume. This suggests that the irreversibility of Hugoniots, i.e., the increase of entropy directly depends on the degree of volume compression. It is consistent with the observations from Ref. [49].

## Appendix A. Derivation of ideal solid *P-V-T* EOS

With pressure and entropy as independent variables, the differential of internal energy becomes:

$$\mathrm{d}U = -P\left(\frac{\partial V}{\partial P}\right)_S \mathrm{d}P + \left[T - P\left(\frac{\partial V}{\partial S}\right)_P\right]\mathrm{d}S \tag{A.1}$$

With the Maxwell relation, $\left(\frac{\partial V}{\partial S}\right)_P = \left(\frac{\partial T}{\partial P}\right)_S$, the second integral factor in equation (A.1) becomes $\left[T - P\left(\frac{\partial T}{\partial P}\right)_S\right]$. The first order derivative of temperature as a function of pressure for equation (2) is:

$$\left(\frac{\partial T}{\partial P}\right)_S = \frac{R_P}{S} \tag{A.2}$$

Then, the second integral factor becomes $\left(T - P\frac{R_P}{S}\right)$ which is equal to zero from equation (2). Therefore, the internal energy of an ideal solid depends on pressure only, $U = U(P)$, whose differential is:

$$\mathrm{d}U = C_S \mathrm{d}P \tag{A.3}$$

where $C_S$ is the specific work at constant entropy. Then, based on the Gibbs equation, $\mathrm{d}U = T\mathrm{d}S - P\mathrm{d}V$, the differential of volume is:

$$\mathrm{d}V = -\frac{C_S}{P}\mathrm{d}P + \frac{T}{P}\mathrm{d}S \tag{A.4}$$

With equation (2), equation (A.4) becomes:

$$\mathrm{d}V = -\frac{C_S}{P}\mathrm{d}P + R_P\frac{T}{P}\mathrm{d}\left(\frac{P}{T}\right) \tag{A.5}$$

Assuming that $C_S$ is constant, the integral of equation (A.5) is:

$$V = -C_S \ln\frac{P}{P_0} + R_P \ln\left(\frac{P}{T}\Big/\frac{P_0}{T_0}\right) + V_0 \tag{A.6}$$

With the relationship, $C_T = C_S - R_P$, the final ideal solid *P-V-T* EOS is obtained:

$$V = -C_T \ln\frac{P}{P_0} - R_P \ln\frac{T}{T_0} + V_0 \tag{A.7}$$

## Appendix B. Symmetric statement of thermodynamic stability

The most common statement of thermodynamic stability is that for an isolated system, entropy will increase and reach maximum value when the system reaches the equilibrium state. This is the principle of entropy increase of isolated system. To show symmetry, thermodynamic stability can be expressed equivalently as follows, i.e., with constant entropy and volume, the internal energy is minimized when the system reaches the equilibrium state. It is mathematically expressed as:

$$\begin{cases} \delta U = 0 \\ \delta^2 U > 0 \end{cases} \tag{B.1}$$

As a function of entropy and volume, the Taylor expansion for internal energy is as follows:

$$\begin{aligned} \mathrm{d}U = \delta U + \delta^2 U + \ldots = \left[\left(\frac{\partial U}{\partial S}\right)_V \mathrm{d}S + \left(\frac{\partial U}{\partial V}\right)_S \mathrm{d}V\right] \\ + \frac{1}{2}\left[\left(\frac{\partial^2 U}{\partial S^2}\right)_V (\mathrm{d}S)^2 + 2\frac{\partial^2 U}{\partial S \partial V}\mathrm{d}S\mathrm{d}V + \left(\frac{\partial^2 U}{\partial V^2}\right)_S (\mathrm{d}V)^2\right] + \ldots \end{aligned} \tag{B.2}$$

Thermodynamic stability requires:

$$\left(\frac{\partial^2 U}{\partial S^2}\right)_V (\mathrm{d}S)^2 + 2\frac{\partial^2 U}{\partial S \partial V}\mathrm{d}S\mathrm{d}V + \left(\frac{\partial^2 U}{\partial V^2}\right)_S (\mathrm{d}V)^2 > 0 \tag{B.3}$$

This requirement can be decomposed into the following three criteria:

$$\begin{cases} \left(\frac{\partial^2 U}{\partial S^2}\right)_V > 0 \\ \left(\frac{\partial^2 U}{\partial V^2}\right)_S > 0 \\ \left(\frac{\partial^2 U}{\partial S^2}\right)_V \left(\frac{\partial^2 U}{\partial V^2}\right)_S - \left(\frac{\partial^2 U}{\partial S \partial V}\right)^2 > 0 \end{cases} \tag{B.4}$$

The first criterion in equation (B.4) is further expressed as:

$$\left(\frac{\partial^2 U}{\partial S^2}\right)_V = \left(\frac{\partial T}{\partial S}\right)_V = \frac{T}{C_V} > 0 \tag{B.5}$$

This criterion states that the temperature must increase when heating at constant volume. Thus, it is called the heat stability criterion.

The second criterion in equation (B.4) is further expressed as:

$$\left(\frac{\partial^2 U}{\partial V^2}\right)_S = -\left(\frac{\partial P}{\partial V}\right)_S = \frac{P}{C_S} > 0 \tag{B.6}$$

This criterion states that the pressure must increase when working at constant entropy. Thus, it can be called the work stability criterion.

The third criterion in equation (B.4) is further expressed as:

$$\left(\frac{\partial^2 U}{\partial S^2}\right)_V \left(\frac{\partial^2 U}{\partial V^2}\right)_S - \left(\frac{\partial^2 U}{\partial S \partial V}\right)^2 = \frac{TP}{C_S C_P} = \frac{TP}{C_V C_T} > 0 \tag{B.7}$$

Combining equation (B.7) with equations (B.5) and (B.6) yields that both the specific heat at constant pressure and the specific work at constant temperature are positive:

$$\begin{cases} C_P > 0 \\ C_T > 0 \end{cases} \tag{B.8}$$

In addition, there is the following relationship in equation (B.7):

$$\frac{C_S}{C_T} = \frac{C_V}{C_P} \tag{B.9}$$

It is deduced from the basic thermodynamic relationship and is therefore always valid. Thermodynamic deduction gives:

$$C_P\left(C_T - C_S\right) = C_T\left(C_P - C_V\right) = TP\left[\left(\frac{\partial S}{\partial P}\right)_T\right]^2 = TP\left[\left(\frac{\partial V}{\partial T}\right)_P\right]^2 > 0 \tag{B.10}$$

Thus, the three criteria of thermodynamic stability can be expressed consistently as:

$$\begin{cases} C_P > C_V > 0 \\ C_T > C_S > 0 \end{cases} \tag{B.11}$$

The third criterion states that when the same amount of heat is transferred into the system, the increase in temperature of isobaric process is smaller than that of isochoric process; when the same amount of work is transferred into the system, the increase in pressure of isothermal process is smaller than that of isentropic process. The above two statements are equivalent.

Further, based on the two relationships, $C_P = C_V + R$ and $C_T = C_S - R_P$, the ideal gas constant must be positive, while the ideal solid constant must be negative:

$$\begin{cases} R > 0 \\ R_P < 0 \end{cases} \tag{B.12}$$

This is what thermodynamic stability requires.

## Appendix C. Derivation of universal EOSs

The *P*-*V*-*T* EOSs are:

$$\begin{cases} V = -C_T \ln\dfrac{P}{P_0} - R_P \ln\dfrac{T}{T_0} + V_0 & \text{for ideal solid limit} \\ PV = RT & \text{for ideal gas limit} \end{cases} \tag{C.1}$$

Their first order derivatives of pressure with respect to volume at constant temperature are:

$$\begin{cases} \left(\dfrac{\partial P}{\partial V}\right)_T = -\dfrac{1}{C_T}\dfrac{1}{\left(\dfrac{1}{P}\right)} & \text{for ideal solid limit} \\ \left(\dfrac{\partial P}{\partial V}\right)_T = -\dfrac{1}{RT}\dfrac{1}{\left(\dfrac{1}{P}\right)^2} & \text{for ideal gas limit} \end{cases} \tag{C.2}$$

Interpolation between these two extremes gives:

$$\left(\frac{\partial P}{\partial V}\right)_T = -\frac{1}{C_T'}\left(\frac{1}{\dfrac{1}{P}} - \frac{1}{\dfrac{1}{P} + \dfrac{C_T'}{R'T}}\right) = -\frac{1}{R'T}\frac{1}{\dfrac{1}{P}\left(\dfrac{1}{P} + \dfrac{C_T'}{R'T}\right)} \tag{C.3}$$

The superscript ' denotes the universal parameters after interpolation. The integral of equation (C.3) with respect to pressure is:

$$V = \frac{R'T}{P} - \frac{R'T}{P_0} - C_T'\ln\frac{P}{P_0} + V\left(T, P_0\right) \tag{C.4}$$

where $V$ ($T$, $P_0$) is an unknown function of temperature. Therefore, another interpolation is needed to relate temperature and volume. The first order derivatives of $1/T$ with respect to volume at constant pressure based on the two parts of equation (C.1) are:

$$\begin{cases} \left(\dfrac{\partial\left(\frac{1}{T}\right)}{\partial V}\right)_P = \dfrac{1}{R_P}\dfrac{1}{T} & \text{for ideal solid limit} \\ \left(\dfrac{\partial\left(\frac{1}{T}\right)}{\partial V}\right)_P = -\dfrac{P}{R}\dfrac{1}{T^2} & \text{for ideal gas limit} \end{cases} \tag{C.5}$$

This is the only relationship between temperature and volume that leads to equations that can be interpolated. Then, interpolation gives:

$$\left(\frac{\partial\left(\frac{1}{T}\right)}{\partial V}\right)_P = \frac{1}{R_P'}\left(\frac{1}{T} - \frac{1}{T - R_P'\frac{P}{R'}}\right) = -\frac{P}{R'}\frac{1}{T\left(T - R_P'\frac{P}{R'}\right)} \tag{C.6}$$

The integral of equation (C.6) with respect to temperature is:

$$V = \frac{R'T}{P} - \frac{R'T_0}{P} - R_P'\ln\frac{T}{T_0} + V\left(T_0, P\right) \tag{C.7}$$

Combining equations (C.4) and (C.7) gives the final universal EOS in *P*-*V*-*T* form:

$$V = R'\left(\frac{T}{P} - \frac{T_0}{P_0}\right) - R_P'\ln\frac{T}{T_0} - C_T'\ln\frac{P}{P_0} + V_0 \tag{C.8}$$

The same methodology is used to derive the universal EOS in *P*-*S*-*T* form. The *P*-*S*-*T* EOSs are:

$$\begin{cases} TS = R_P P & \text{for ideal solid limit} \\ S = C_P\ln\dfrac{T}{T_0} - R\ln\dfrac{P}{P_0} + S_0 & \text{for ideal gas limit} \end{cases} \tag{C.9}$$

Their first order derivatives of temperature with respect to entropy at constant pressure are:

$$\begin{cases} \left(\frac{\partial T}{\partial S}\right)_P = -\frac{1}{R_P P}\frac{1}{\left(\frac{1}{T}\right)^2} & \text{for ideal solid limit} \\ \left(\frac{\partial T}{\partial S}\right)_P = \frac{1}{C_P}\frac{1}{\left(\frac{1}{T}\right)} & \text{for ideal gas limit} \end{cases} \quad (C.10)$$

Then, interpolation of these two equations gives:

$$\left(\frac{\partial T}{\partial S}\right)_P = \frac{1}{C_P'}\left(\frac{1}{\frac{1}{T}} - \frac{1}{\frac{1}{T} - \frac{C_P'}{R_P'P}}\right) = -\frac{1}{R_P'P}\frac{1}{\frac{1}{T}\left(\frac{1}{T} - \frac{C_P'}{R_P'P}\right)} \quad (C.11)$$

The integral of equation (C.11) with respect to temperature gives:

$$S = \frac{R_P'P}{T} - \frac{R_P'P}{T_0} + C_P'\ln\frac{T}{T_0} + S(T_0, P) \quad (C.12)$$

The first order derivatives of $1/P$ with respect to entropy at constant temperature based on the two parts of equation (C.9) are:

$$\begin{cases} \left(\frac{\partial\left(\frac{1}{P}\right)}{\partial S}\right)_T = -\frac{T}{R_P}\frac{1}{P^2} & \text{for ideal solid limit} \\ \left(\frac{\partial\left(\frac{1}{P}\right)}{\partial S}\right)_T = \frac{1}{R}\frac{1}{P} & \text{for ideal gas limit} \end{cases} \quad (C.13)$$

Interpolation between these two equations gives:

$$\left(\frac{\partial\left(\frac{1}{P}\right)}{\partial S}\right)_T = \frac{1}{R'}\left(\frac{1}{P} - \frac{1}{P - R'\frac{T}{R_P'}}\right) = -\frac{T}{R_P'}\frac{1}{P\left(P - R'\frac{T}{R_P'}\right)} \quad (C.14)$$

The integral of equation (C.14) with respect to pressure gives:

$$S = \frac{R_P'P}{T} - \frac{R_P'P_0}{T} - R'\ln\frac{P}{P_0} + S(T, P_0) \quad (C.15)$$

Combining equations (C.12) and (C.15) gives the final universal EOS in *P*-*S*-*T* form:

$$S = R_P'\left(\frac{P}{T} - \frac{P_0}{T_0}\right) - R'\ln\frac{P}{P_0} + C_P'\ln\frac{T}{T_0} + S_0 \quad (C.16)$$